\documentclass[prc,aps,twocolumn,showpacs,superscriptaddress]{revtex4} 
\usepackage{graphicx}
\usepackage[utf8]{inputenc}
\usepackage{dcolumn}
\begin{document}
\pagebreak
\title{Shape coexistence at the proton drip-line: First identification of excited states in $^{180}$Pb}
\author{P.~Rahkila}
\email{panu.rahkila@phys.jyu.fi}
\affiliation{Department of Physics, University of Jyv\"askyl\"a, FI-40014 University of Jyv\"askyl\"a,  Finland}
\author{D.G.~Jenkins}
\affiliation{Department of Physics, University of York, Heslington, York YO10 5DD, United Kingdom}
\author{J.~Pakarinen}
\altaffiliation[Present address:~]{ISOLDE, CERN, CH-1211 Geneva, Switzerland}
\affiliation{Oliver Lodge Laboratory, University of Liverpool, Liverpool L69 7ZE, United Kingdom}
\author{C. Gray-Jones}
\affiliation{Oliver Lodge Laboratory, University of Liverpool, Liverpool L69 7ZE, United Kingdom}
\author{P.T.~Greenlees}
\affiliation{Department of Physics, University of Jyv\"askyl\"a, FI-40014 University of Jyv\"askyl\"a,  Finland}
\author{U.~Jakobsson}
\affiliation{Department of Physics, University of Jyv\"askyl\"a, FI-40014 University of Jyv\"askyl\"a,  Finland}
\author{P.~Jones}
\affiliation{Department of Physics, University of Jyv\"askyl\"a, FI-40014 University of Jyv\"askyl\"a,  Finland}
\author{R.~Julin}
\affiliation{Department of Physics, University of Jyv\"askyl\"a, FI-40014 University of Jyv\"askyl\"a,  Finland}
\author{S.~Juutinen}
\affiliation{Department of Physics, University of Jyv\"askyl\"a, FI-40014 University of Jyv\"askyl\"a,  Finland}
\author{S.~Ketelhut}
\affiliation{Department of Physics, University of Jyv\"askyl\"a, FI-40014 University of Jyv\"askyl\"a,  Finland}
\author{H.~Koivisto}
\affiliation{Department of Physics, University of Jyv\"askyl\"a, FI-40014 University of Jyv\"askyl\"a,  Finland}
\author{M.~Leino}
\affiliation{Department of Physics, University of Jyv\"askyl\"a, FI-40014 University of Jyv\"askyl\"a,  Finland}
\author{P.~Nieminen}
\affiliation{Department of Physics, University of Jyv\"askyl\"a, FI-40014 University of Jyv\"askyl\"a,  Finland}
\author{M.~Nyman}
\altaffiliation[Present address:~]{Laboratory of Radiochemistry, University of Helsinki, FI-00014 University of Helsinki, Finland}
\affiliation{Department of Physics, University of Jyv\"askyl\"a, FI-40014 University of Jyv\"askyl\"a,  Finland}
\author{P.~Papadakis}
\affiliation{Oliver Lodge Laboratory, University of Liverpool, Liverpool L69 7ZE, United Kingdom}
\author{S.~Paschalis}
\altaffiliation[Present address:~]{Nuclear Science Division, Lawrence Berkeley National Laboratory, Berkeley, CA 94720, USA}
\affiliation{Oliver Lodge Laboratory, University of Liverpool, Liverpool L69 7ZE, United Kingdom}
\author{M.~Petri}
\altaffiliation[Present address:~]{Nuclear Science Division, Lawrence Berkeley National Laboratory, Berkeley, CA 94720, USA}
\affiliation{Oliver Lodge Laboratory, University of Liverpool, Liverpool L69 7ZE, United Kingdom}
\author{P.~Peura}
\affiliation{Department of Physics, University of Jyv\"askyl\"a, FI-40014 University of Jyv\"askyl\"a,  Finland}
\author{O.J.~Roberts}
\affiliation{Department of Physics, University of York, Heslington, York YO10 5DD, United Kingdom}
\author{T.~Ropponen}
\altaffiliation[Present address:~]{National Superconducting Cyclotron Laboratory, Michigan State University, East Lansing, MI 48824, USA}
\affiliation{Department of Physics, University of Jyv\"askyl\"a, FI-40014 University of Jyv\"askyl\"a,  Finland}
\author{P.~Ruotsalainen}
\affiliation{Department of Physics, University of Jyv\"askyl\"a, FI-40014 University of Jyv\"askyl\"a,  Finland}
\author{J.~Sarén}
\affiliation{Department of Physics, University of Jyv\"askyl\"a, FI-40014 University of Jyv\"askyl\"a,  Finland}
\author{C.~Scholey}
\affiliation{Department of Physics, University of Jyv\"askyl\"a, FI-40014 University of Jyv\"askyl\"a,  Finland}
\author{J.~Sorri}
\affiliation{Department of Physics, University of Jyv\"askyl\"a, FI-40014 University of Jyv\"askyl\"a,  Finland}
\author{A.G.~Tuff}
\affiliation{Department of Physics, University of York, Heslington, York YO10 5DD, United Kingdom}
\author{J.~Uusitalo}
\affiliation{Department of Physics, University of Jyv\"askyl\"a, FI-40014 University of Jyv\"askyl\"a,  Finland}
\author{R.~Wadsworth}
\affiliation{Department of Physics, University of York, Heslington, York YO10 5DD, United Kingdom}
\author{M.~Bender}
\affiliation{Universit\'e Bordeaux, CNRS/IN2P3, Centre d'Etudes Nucl{\'e}aires de Bordeaux Gradignan, CENBG, Chemin du Solarium, BP120, F-33175 Gradignan, France}
\author{P.-H.~Heenen}
\affiliation{Service de Physique Nucl\'eaire Théorique, Universit\'e Libre de Bruxelles, B-1050 Bruxelles, Belgium}

\date{\today}

\begin{abstract}

Excited states in the extremely neutron-deficient nucleus, $^{180}$Pb, have been identified for the first time using the JUROGAM II array in conjunction with the RITU recoil separator at the Accelerator Laboratory of the University of Jyv\"askyl\"a. This study lies at the limit of what is presently achievable with in-beam spectroscopy, with an estimated cross-section of only 10~nb for the $^{92}$Mo($^{90}$Zr,2n)$^{180}$Pb reaction. A continuation of the trend observed in $^{182}$Pb and $^{184}$Pb is seen, where the prolate minimum continues to rise beyond the $N=104$ mid-shell with respect to the spherical ground state. Beyond mean-field calculations are in reasonable correspondence with the trends deduced from experiment.

\end{abstract}

\pacs{21.60.Ev,23.20.Lv,27.80.+w}    
\maketitle

Descriptions of the atomic nucleus imply complementarity between single-particle structure and collective phenomena. The interplay between these two aspects can lead to the nucleus being driven, through structural effects, to adopt different mean-field shapes for a small cost in energy - a phenomenon frequently described as nuclear shape coexistence. The light lead nuclei have long been highlighted as a dramatic example of such shape coexistence with compelling evidence in favor of this picture coming from $\alpha$-decay studies. For example, fine structure is observed in the $\alpha$ decay of $^{190}$Po which feeds two excited 0$^{+}$ states as well as the ground state of $^{186}$Pb~\cite{andreyev}. Hindrance factors for the three $\alpha$ branches support a picture where three shape minima: prolate, oblate and spherical co-exist within a narrow range of excitation energy. A complementary strand has been to locate excited states in the very neutron-deficient lead nuclei via in-beam spectroscopy. This approach is very challenging given the very small production cross-sections involved and the overwhelming background stemming from fission products. The relevant experimental technique here is recoil-decay tagging (RDT)~\cite{schmidt,simon,paul} where recoiling residues, typically produced in a heavy-ion fusion evaporation reaction, are separated from beam-like particles and fission products using a recoil separator, and implanted into a position-sensitive silicon detector at the focal plane. Gamma rays observed in an array surrounding the target position may then be correlated with the characteristic decay of these exotic nuclei detected in the silicon detector, closely following the implantation of the respective recoil. In such a manner, it was possible to study the excited states of $^{184}$Pb~\cite{cocks} and $^{182}$Pb~\cite{jenkins} at the University of Jyv\"askyl\"a (JYFL) in the late 1990s. In the latter example, the production cross-section for $^{182}$Pb was only $\sim$300~nb, corresponding to the veritable needle-in-the-haystack. The studies of $^{182}$Pb and $^{184}$Pb, taken together, confirmed that the prolate configuration reaches an energetic minimum at the neutron mid-shell (N=104); a review of this work can be found in ref.~\cite{julin}. 

In recent years, the focus has switched from measurements at the limits of experimental sensitivity, to more detailed spectroscopic measurements and high-spin studies. For example, Dracoulis {\em et al.} have carried out studies revealing the details of high-lying isomers in lead nuclei around $A=190$~\cite{dracoulis1,dracoulis2,dracoulis3,dracoulis4}, and Pakarinen~{\em et al.} have determined the position of the excited band in $^{186}$Pb~\cite{pakarinen1,pakarinen2} believed to be associated with the oblate configuration. A comprehensive picture of the evolution of collectivity in these nuclei, however, can only come from determination of electromagnetic matrix elements. This task has begun through a determination of transition matrix elements in $^{186}$Pb and $^{188}$Pb via lifetime measurements using a plunger device - the so-called recoil distance method~\cite{dewald,grahn1,grahn2}. In the future, transition {\em and} diagonal matrix elements, the latter being uniquely sensitive to the sign of the nuclear deformation, may be obtained using Coulomb excitation of radioactive ion beams. Such pioneering studies in the Z$\simeq$82 region have recently been carried out for the light mercury and radon nuclei at the REX-ISOLDE facility in CERN. 

A further degree of freedom which has yet to be fully explored in the light lead nuclei is the polarisation of the nuclear shape from the addition of an extra neutron. Attempts to study odd-A light lead nuclei have proven very difficult given the small cross-sections, and complex level schemes which are hard to disentangle with the low statistics of available $\gamma-\gamma$ coincidences and a high proportion of converted transitions. Despite these difficulties, the first results on odd-A lead nuclei below the mid-shell were recently reported by Pakarinen {\em et al.}~\cite{pakarinen3}, and support the dominance of a prolate shape in $^{185}$Pb. The recently commissioned SAGE spectrometer at JYFL~\cite{sage} promises to provide the unique possibility of measuring conversion electron-$\gamma$ ray coincidences correlated with exotic nuclei using the recoil-decay-tagging technique. This will open up the possibilities for studies of the odd-A nuclei.

Parallel to these exciting new directions outlined above, our intention in the present work was to extend our understanding of the even-even light lead nuclei to the extremes of neutron-deficiency, namely to $^{180}$Pb. Recent beyond mean field calculations~\cite{bender1,egido,rodriguez} extending down to $A=182$ predict the disappearance of the oblate minimum in the lightest lead isotopes. As has already been established, the prolate minimum in the light lead nuclei reaches its lowest excitation energy at the mid-shell ($N=104$) and is expected to rise rather rapidly in the lighter nuclei. By moving to $^{180}$Pb, we are truly on the edge of the proton drip-line; the most recent AME mass evaluation lists one proton ($S(p)$) and two proton ($S(2p)$) separation energies for $^{180}$Pb of 930(50)~keV and 200(25)~keV, respectively~\cite{ame2003-2}. 

A range of different fusion-evaporation reactions can be employed in studying the light lead nuclei. In the past, asymmetric reactions such as $^{144}$Sm($^{42}$Ca,4n)$^{182}$Pb were used for RDT studies at JYFL. An attractive alternative is to use symmetric cold-fusion reactions. Keller {\em et al.} studied a number of such reactions including $^{90}$Zr+$^{90}$Zr~\cite{keller}. These reactions have excellent characteristics for a RDT measurement as the number of open evaporation channels is small and fission survivability is greatly improved relative to a typical asymmetric reaction. In fact, the very first RDT measurement was carried out with the $^{90}$Zr+$^{90}$Zr reaction at GSI in the 1980s~\cite{schmidt,simon}. For $^{180}$Pb, in particular, these type of Zr- or Mo-induced reactions are rather favourable and have been employed in the past, for example by Toth {\em et al.}, who used the $^{92}$Mo($^{90}$Zr,2n)$^{180}$Pb reaction at 420~MeV to study $^{180}$Pb using the FMA recoil separator~\cite{toth2}, after initially identifying it at LBNL~\cite{toth1}. With a gas-filled separator such as the RITU separator at JYFL~\cite{ritu}, the transmission efficiency is well over 50\% for such symmetric reactions~\cite{saren}. 

In the past, when the light lead isotopes were studied at JYFL, zirconium and molybdenum beams were unavailable. Such beams have now become available in the last year, following development of a sputtering method with the JYFL ECR ion source which can generate the beam currents needed for in-beam experiments. Given the advantages of using a Zr-induced reaction, as described above, we chose to exploit these recent developments in this present measurement and this contributed significantly to our ability to successfully study a nucleus as exotic as $^{180}$Pb. Indeed, this beam development opens up important additional opportunities to study neighbouring nuclei in a more sensitive and effective manner.

A $^{90}$Zr beam from the K130 cyclotron at JYFL was accelerated to 400 MeV and was incident on a metallic, self-supporting 1~mg/cm$^2$ target of $^{92}$Mo isotopically enriched to 96\%. A 50 $\mu$g/cm$^2$ carbon reset foil was placed directly behind the target. Prompt $\gamma$ rays were detected with the JUROGAM II array comprising 24 EUROGAM clover detectors \cite{clover} and 15 EUROGAM phase one \cite{phase1} or GASP \cite{gasp} detectors. The evaporation residues were separated from fission products and beam particles using the RITU recoil separator \cite{ritu} and were implanted in a 300~$\mu$m-thick double-sided silicon strip detector (DSSSD) which comprises part of the GREAT focal-plane spectrometer \cite{great}. The triggerless Total Data Readout (TDR) data acquisition \cite{tdr} was used to collect the data. Digital TNT2 electronics~\cite{tnt2} were used to instrument the clover detectors of the JUROGAM II array. Data from the two sources were merged in the TDR software and the Grain software package~\cite{grain} was used to construct the events and to analyse the data.  

\begin{figure*}[htb]
\includegraphics[angle=0,width=\linewidth]{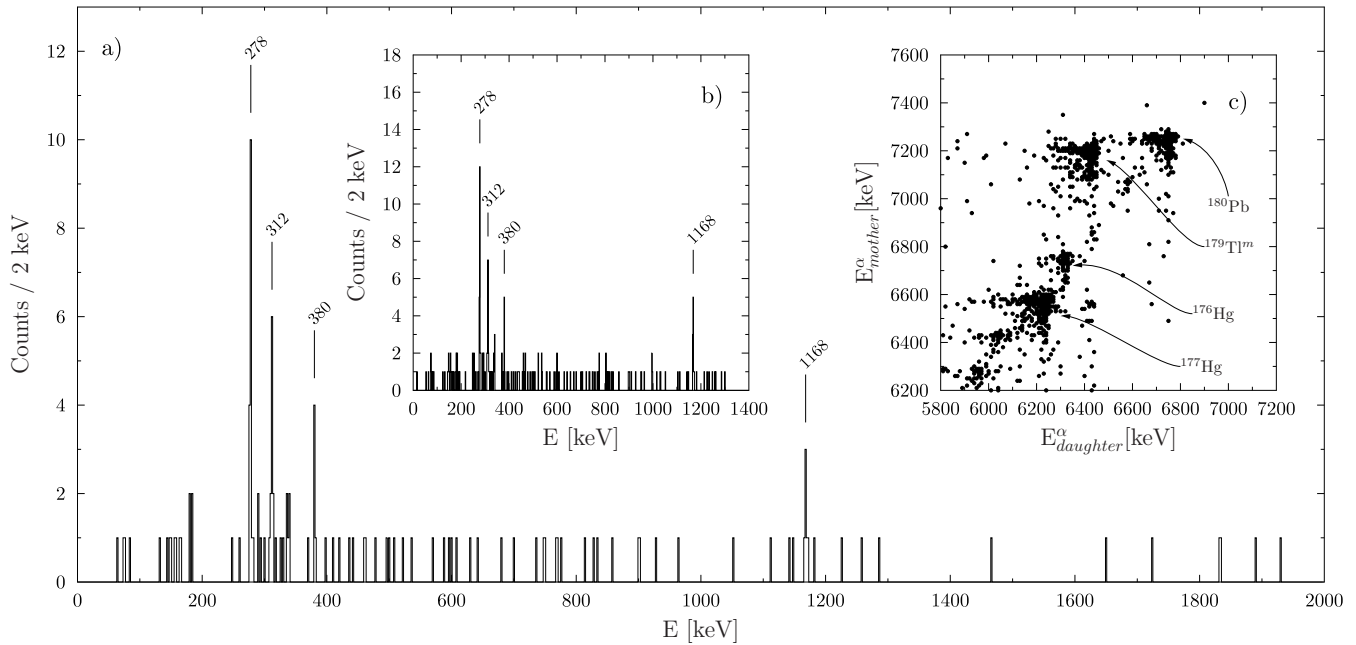}
\caption{Singles $\gamma$-ray energy spectra tagged by genetic correlations of the $^{180}$Pb decay chains. Panel a) shows a spectrum generated by demanding a full energy recoil-$\alpha^F$-$\alpha^F$ chain as a tag, whereas for panel b) recoil-$\alpha^F$/$\alpha^E$-$\alpha^E$/$\alpha^F$-$\alpha^F$ chains were also accepted. See text for details. In panel c) correlation plot showing the daughter $\alpha$-particle energy vs. mother $\alpha$ particle-energy is presented. The correlations were made with a search time of 12~ms for the recoil-mother pair and 60~ms for the mother-daughter pair. The correlated mother-daughter decays are indicated according to the respective mother nucleus.}
\label{fig:gammas}       
\end{figure*}

The RDT analysis was complicated in the present study by the presence of a 1.5~ms $\alpha$-decaying isomer in $^{179}$Tl~\cite{nds179}. Due to the finite resolution of the silicon detectors, the peak corresponding to the $\alpha$ decay of $^{179}$Tl$^m$ partially overlaps the peak corresponding to $^{180}$Pb (see Fig.~\ref{fig:gammas}) and thus prevents the standard approach of selecting the channel by gating on $\alpha$ particles from the decay of the nucleus of interest, i.e. $^{180}$Pb. Fortunately, both $^{176}$Hg the $\alpha$-decay daughter of $^{180}$Pb, and the grand-daughter, $^{172}$Pt, have rather short half-lives (20~ms~\cite{nds176} and 96~ms~\cite{nds172}, respectively) and large $\alpha$-decay branches ($\sim$95\% in both cases \cite{nds176,nds172}) allowing the use of multi-step genetic correlations to be used to cleanly tag $^{180}$Pb recoils. The cleanest manner of tagging, which is free of ambiguity, is to use only the full-energy $\alpha$ ($\alpha^F$) correlations of $^{180}$Pb and $^{176}$Hg (see Fig.~\ref{fig:gammas}) but this will not recover all of the events which may be correlated, in principle, since about 40\% of the emitted $\alpha$ particles escape the implantation detector without depositing their full energy ($\alpha^E$). A significant fraction of the lost events can be salvaged by exploiting the fact that both $^{180}$Pb and $^{176}$Hg have rather short half-lives, by allowing $\alpha$ particles from either $^{180}$Pb or $^{176}$Hg to escape and requiring that these recoil-$\alpha^F$-$\alpha^E$ or recoil-$\alpha^E$-$\alpha^F$ chains are followed by a $^{172}$Pt full-energy $\alpha$-decay. 
 
The $\alpha$-decay energy and half-life of $^{180}$Pb were extracted from recoil-$^{180}$Pb-$^{176}$Hg correlations.  The decay event spectra were calibrated internally using the known $\alpha$ activities of $^{180,179,177}$Hg and $^{176}$Pt \cite{ensdf}. The value of 7254(7)~keV obtained for the $^{180}$Pb $\alpha$-decay energy and 4.1(3)~ms for the half-life are in good correspondence with recent measurements by Andreyev {\em et al.}~\cite{andrei180}.

In total, 271 full energy recoil-$\alpha^F$-$\alpha^F$ correlations were observed during the 160~h of irradiation with an average beam current of 7~pnA. Taking into account the measured 60\% probability for observing the full energy of the $\alpha$ particles and the assumed 80\% coverage of the focal plane distribution and 60\% RITU transmission, the production cross section for $^{180}$Pb is estimated to be 10~nb, one of the lowest cross-sections ever exploited in a successful in-beam spectroscopy experiment.

The singles $\gamma$-ray spectra correlated  with the recoil-$\alpha^F$-$\alpha^F$ and the recoil-$\alpha^F$/$\alpha^E$-$\alpha^E$/$\alpha^F$-$\alpha^F$ decay chains are presented in Fig.~\ref{fig:gammas}. The recoil-$\alpha^F$-$\alpha^F$ spectrum was used to unambiguously identify the $\gamma$-rays in $^{180}$Pb. Four $\gamma$-ray transitions with energies of 278, 312, 380 and 1168~keV are firmly assigned to $^{180}$Pb. The recoil-$\alpha^F$/$\alpha^E$-$\alpha^E$/$\alpha^F$-$\alpha^F$ correlated spectrum, which shares the same features as the recoil-$\alpha^F$-$\alpha^F$ spectrum, is used to extract the intensities of the transitions since at this level of statistics, the statistical errors are significant. The details are presented in table~\ref{tab:gammas}. 

\begin{table}[htb]
\caption{Gamma-ray transitions assigned to $^{180}$Pb in the present work. The energies (E$_\gamma$) in keV, raw intensities (counts), relative intensities without (I$_{rel}$) and with the correction for internal conversion (I$_{rel,icc}$) assuming pure E2 character, as well as the tentative level assignments are given.}
\label{tab:gammas}
\begin{ruledtabular}
\begin{tabular}{D{-}{(}{-1}D{-}{(}{-1}D{-}{(}{-1}D{-}{(}{-1}D{-}{\rightarrow}{-1}}
\multicolumn{1}{c}{$E_\gamma$ (keV)}&  
\multicolumn{1}{c}{counts} &
\multicolumn{1}{c}{$I_{rel}$ (\%)}&
\multicolumn{1}{c}{$I_{rel,icc}$ (\%)}&
I^\pi_i - I^\pi_f\\
\hline
278 -1) & 19 -5) & 93 -21) & 107 -25) & \left(4^+\right) - \left(2^+\right)\\
312 -1) & 11 -4) & 56 -17) & 62 -19) & \left(6^+\right) - \left(4^+\right)\\
380 -1) & 6 -3)  & 33 -14) & 36 -15) & \left(8^+\right) - \left(6^+\right)\\
1168 -1) & 9 -3)  & 100 -33) & 100 -33) & \left(2^+\right) - ~~0^+\\
\end{tabular}
\end{ruledtabular}
\end{table}

\begin{figure}[htb]
\includegraphics[angle=0,width=\linewidth]{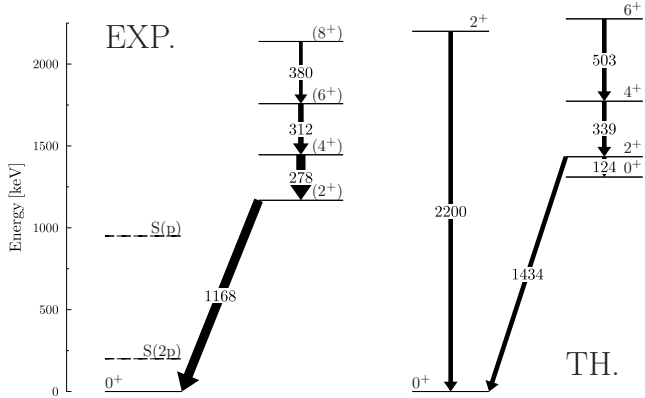}
\caption{Comparison between the layout of excited states proposed from the experimental data and that suggested by beyond mean-field calculations (see text). The width of the arrows denotes the measured intensities for the experimental data, while carrying no meaning for the theoretical predictions. The dashed lines give the location of the one-proton, S(p), and two-proton, S(2p), separation energies~\cite{ame2003-2}.}
\label{fig:comparison}       
\end{figure}

Assuming that the observed $\gamma$ rays form a cascade, then they can be ordered on the basis of their relative intensities and on systematics, since statistics are too low to obtain $\gamma$-$\gamma$ coincidences or to allow an angular distribution or correlation analysis. Accordingly, it is proposed that the 1168-keV $\gamma$ ray, which has a much higher energy than the other $\gamma$ rays, corresponds to the 2$^{+}$ $\rightarrow$ 0$^{+}$ transition in keeping with that observed in $^{182-188}$Pb, while the three further $\gamma$ rays belong to a rotational band above the 2$^{+}$ state. The proposed level scheme is presented in Fig. \ref{fig:comparison}. It should be noted that all excited states are ``unbound'' in the sense that they lie above the one- and two-proton separation energies~\cite{ame2003-2}.

\begin{figure}[htb]
\includegraphics[angle=0,width=\linewidth]{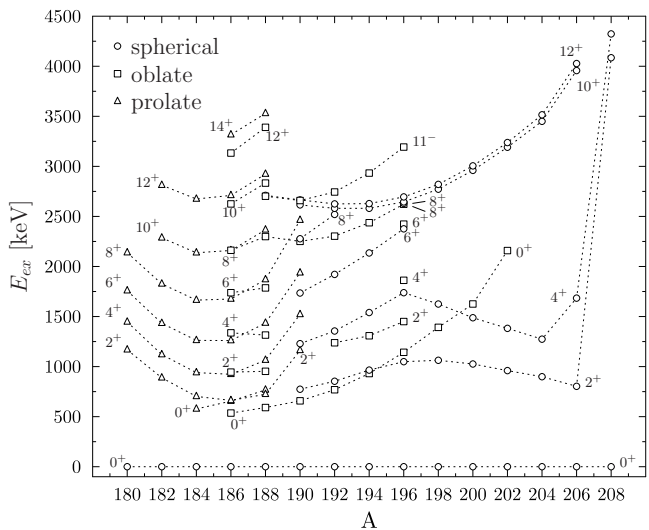}
\caption{Energy level systematics of lead nuclei with A$\leq$208. The data are taken from the present work, ref.~\cite{jenkins}, ref.~\cite{ensdf}, and references therein.}
\label{fig:systematics}       
\end{figure}

The proposed level scheme is in good agreement, in a qualitative sense, with a smooth extrapolation of the trends seen in the neighbouring even-even lead nuclei q.v. Fig.~\ref{fig:systematics}, where the prolate-deformed states are suggested to move up in energy relative to the ground state beyond the neutron mid-shell at $N=104$.

The same beyond mean-field method as is described in ref.~\cite{bender1} has been used to calculate the structure of $^{180}$Pb (see fig.~\ref{fig:comparison}).  This method is based on the configuration mixing of self-consistent mean-field wave functions. After projection on angular momentum and particle number, states with different intrinsic axial quadrupole moment are mixed within the Generator Coordinate Method. The final wave functions are usually spread over a wide range of intrinsic deformations. Within this approach it is possible to calculate the spectrum associated with the axial collective mode and the electromagnetic transition probabilities allowed between excited states.  The Skyrme interaction SLy6 and a density-dependent pairing interaction have been used. An extensive description of the method can be found in refs.~\cite{grahn2,pakarinen3,bender1}. 

As for heavier Pb isotopes, the calculated ground state of $^{180}$Pb is dominated by configurations with small deformation close to sphericity. The collective wave function of the excited $2^+$ level at 2.2~MeV suggests its interpretation as a vibrational state. It decays by a strong E2 transition to the ground state. The first excited $0^+$ is predominantly composed of projected prolate mean-field configurations and is the bandhead of a rotational band. The transition quadrupole moments within this band vary between 900 and 950 $fm^2$, which can be translated into a $\beta_{2}$ value (see ref.~\cite{bender1} for its definition) between 0.3 and 0.32, slightly larger than for heavier Pb isotopes. The de-excitation probability of the first $2^+$ state to the ground state is much smaller, with a transition quadrupole moment of 90~$fm^2$. This picture of $^{180}$Pb supports the interpretation of the experimental spectrum (see fig.~\ref{fig:comparison}). The excitation energies of excited states is overestimated by the calculations in comparison to the data. This deficiency has already been noted in earlier calculations for this mass region~\cite{grahn2,pakarinen3,bender1,egido,rodriguez}, and may relate to the absence of time-reversal symmetry breaking in the mean-field wave functions. Breaking this symmetry would decrease the energies of the states with $J \neq 0$ but would not affect the ground state.

\begin{figure}[!tb]
\includegraphics[angle=0,width=\linewidth]{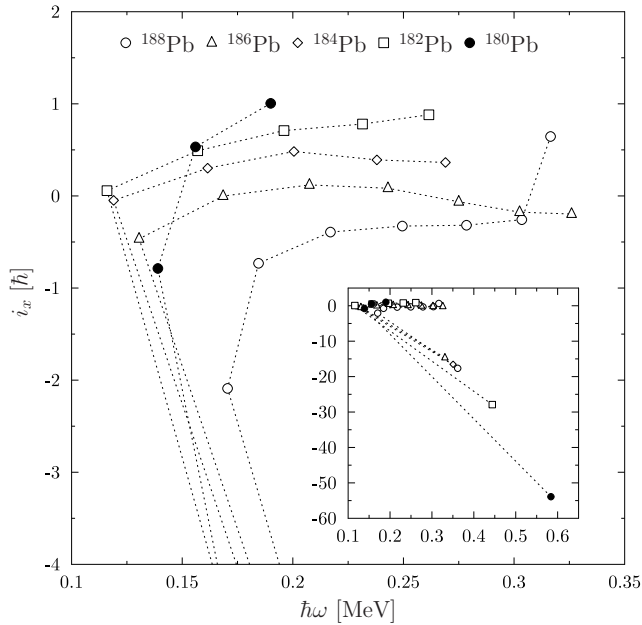}
\caption{Aligned angular momentum $i_x$ of the prolate bands in light Pb isotopes. The insert shows the full scale plot with the 2$^+$ $\rightarrow$ 0$^+$ transitions included. An appropriate Harris reference ${\mathcal J}_0 = 27{\hbar^2}/{MeV}$, ${\mathcal J}_1 = 199{\hbar^4}/{MeV^3}$ has been subtracted.}
\label{fig:moms}       
\end{figure}

The aligned angular momenta for the proposed prolate bands in the light lead nuclei are shown in Fig.~\ref{fig:moms}. At this level of detail, differences are seen between the behavior of $^{180}$Pb and its neighbours, $^{182}$Pb and $^{184}$Pb. In particular, the 4$^{+}$ $\rightarrow$ 2$^{+}$ transition energy is almost 50~keV larger than that for $^{182}$Pb, while the 6$^{+}$ $\rightarrow$ 4$^{+}$ transition energies are almost identical in the two nuclei. In the case of $^{182}$Pb, it was concluded from a variable moment-of-inertia fit to the states above $J=6$, that the 4$^{+}$ state is depressed slightly from its expected location by a few keV but that the 2$^{+}$ state is more strongly depressed in energy, presumably due to mixing ~\cite{jenkins}. Extending these conclusions would imply that the 2$^{+}$ state in $^{180}$Pb is even more strongly depressed from its expected location. Such behavior is also seen in the low-energy part of the prolate band in $^{188}$Pb (see Fig.~\ref{fig:moms}) where it has been attributed to strong mixing with spherical and oblate configurations co-existing at similar energies \cite{dracoulis2,dewald,grahn1,grahn2}. While there is superficial similarity between the low energy behaviour of the prolate band in $^{180}$Pb and the one in $^{188}$Pb, the origin is likely to be different since in $^{180}$Pb the oblate structure is predicted to disappear or lie at much higher energy. This suggests that it may be mixing with the spherical 2$^{+}$ state which is a more likely explanation in the case of $^{180}$Pb. 
 
In summary, excited states in $^{180}$Pb have been identified for the first time using the recoil-decay tagging technique. The cross-section for producing this near-dripline nucleus is only 10~nb, which is on the absolute limit of experimental sensitivity at the present time. Nevertheless, it was possible to identify four $\gamma$ rays which we suggest form a cascade connecting the yrast states. The implied level energies are in good agreement with a smooth extrapolation of systematics obtained for neighbouring nuclei and are in good agreement with the results of beyond mean-field calculations.

\begin{acknowledgments}
This work has been supported by the EU 6th Framework programme ``Integrating Infrastructure Initiative - Transnational Access'', Contract Number: RII3-CT-2004-506065 (EURONS), the European Research Council under the European Community's 7th Framework Programme (FP7/2007-2013) / ERC grant agreement number 203481, by the Academy of Finland under the Finnish Centre of Excellence Programme 2006-2011 (Nuclear and Accelerator Based Physics Programme at JYFL), Contract Number: 213503, by the UK Science and Technology Facilities Council (STFC) and by the IUP-Belgian State Science Policy, BriX newtwork P6/23. PN (Contract Number: 121110) and PTG (Contract Number: 111965) acknowledge the support of the Academy of Finland. We thank the GAMMAPOOL European Spectroscopy Resource for the loan of Germanium detectors for JUROGAM II, and the French CNRS/IN2P3 for the use of the TNT2-D digital electronics. Valuable discussions with Kris Heyde, John Wood and Mikael Sandzelius are gratefully acknowledged.
\end{acknowledgments}

\end{document}